\documentclass[aps,showpacs]{revtex4}

\usepackage{amsmath}
\usepackage{bm}
\usepackage{txfonts}
\usepackage[english]{babel}
\usepackage{subfigure}
\usepackage{graphicx}

\begin{document}

\title{Spin kinetic theory --- quantum kinetic theory in extended phase space}

\author{Mattias Marklund} 
\affiliation{Department of Physics, Ume{\aa} University, SE--901 87 Ume{\aa}, Sweden}

\author{Jens Zamanian} 
\affiliation{Department of Physics, Ume{\aa} University, SE--901 87 Ume{\aa}, Sweden}

\author{Gert Brodin} 
\affiliation{Department of Physics, Ume{\aa} University, SE--901 87 Ume{\aa}, Sweden}

\begin{abstract}
The concept of phase space distribution functions and their evolution is used in the case of en enlarged phase space. In particular, we include the intrinsic spin of particles and present a quantum kinetic evolution equation for a scalar quasi-distribution function. In contrast to the proper Wigner transformation technique, for which we expect the corresponding quasi-distribution function to be a complex matrix, we introduce a spin projection operator for the density matrix in order to obtain the aforementioned scalar quasi-distribution function. There is a close correspondence between this projection operator and the Husimi (or Q) function used extensively in quantum optics. Such a function is based on a Gaussian smearing of a Wigner function, giving a positive definite distribution function. Thus, our approach gives a Wigner-Husimi quasi-distribution function in extended phase space, for which the reduced distribution function on the Bloch sphere is strictly positive. We also discuss the gauge issue and the fluid moment hierarchy based on such a quantum kinetic theory. 
\end{abstract}

\maketitle

\section{Introduction}

In 1932, Eugene Wigner published the paper \textit{On the Quantum Correction For Thermodynamic Equilibrium} \cite{wigner} in which he set out to define the concept of a (quasi-)distribution function in order to approach the problem indicated in the title of his work. This seminal paper was very much the starting point for quantum kinetic theory (see also \cite{weyl,groenewold,moyal}), in which the classical dynamical theory of Boltzmann was extended to the quantum domain. While the Wigner function is of interest in e.g. the interpretation of quantum mechanics \cite{zachos,kurtsiefer}, the introduction of the quasi-distribution function (so called because of not being positive definite) of Wigner is not unique. In particular, within the field of quantum optics several other definitions naturally occur (see e.g. \cite{Leonhardt1997} for an overview). These different definitions correspond to particular operator ordering and thus certain applications lends themselves naturally to different definitions. For example, when considering optical coherence normally ordered operators occurs naturally and hence the  Glauber-Sudarshan P-distribution \cite{Glauber1963,Sudarshan1963} is a convenient choice. On the other hand, anti-normal ordered operators i.e. the Q-function or the more general Husimi function \cite{Husimi1940} are useful when dealing with quantum chaotic systems. For reviews of the subject see for example Refs.\ \cite{zachos,hillery,lee}. Apart from the fields of plasma physics and quantum optics, where different phase space models have been applied with great success, the field of condensed matter physics and transport theory has also utilized the different kinetic(-like) approaches, such as semiclassical techniques \cite{haug-jauho}, Green's function techniques \cite{baym-kadanoff,kadanoff-baym,keldysh1,keldysh2}, and diagrammatic techniques \cite{rammer}. The research field known as quantum plasmas has recently become an area of intense investigations, with possible application to high intensity laser-plasma interactions \cite{eliezer,glenzer,kritcher,lee-etal}, high energy density physics \cite{drake}, , nano- and sub micron-technology, e.g.\ quantum dots \cite{alivisatos,haas,manfredi-hervieux} and plasmonic components \cite{maier,epl}, and nonlinear collective quantum problems \cite{anderson-etal:2002,marklund:2005,shukla-etal:2006,nshukla:2009,shukla-eliasson2009,Brodin-Marklund-Manfredi,Zamanian-Marklund-Brodin}.

The above mentioned studies often assume an unmagnetized state of the system \cite{melrose}. However, significant influence can be exerted by external magnetic fields under the right conditions. For example, the magnetization dynamics of materials have important applications to data storage and memory production \cite{stancil-prabhakar}. Such dynamics is traditionally approached using the Landau-Lifshitz-Gilbert equation \cite{bertotti-etal}. Due to the large set of possible applications, as well as an interest to understand the basic dynamics of such systems, quantum plasma systems with spin has thus recently attracted a lot of interest from the research community (see e.g. \cite{marklund-brodin2007,shukla-eliasson2009} and references therein). In the nonlinear regime spin solitons \cite{brodin-marklund_pop:2007} and ferromagnetic behavior in plasmas can be found \cite{brodin-marklund_pre:2007}. Up until this date, most of the presented studies have been of a theoretical nature. However, it is not difficult to envision future applications to e.g. plasmonic devices \cite{maier} or femtosecond physics \cite{grossman}. 

In this paper we will give an overview of the field of spin extended phase space, and their implications for the formulation of a generalized kinetic theory.

\section{Nonrelativistic microscopic equations: Schr\"odinger and Pauli dynamics}

\subsection{The Schr\"odinger description}

The dynamics of an nonrelativistic scalar electron, represented by its wave function $\psi$, in an external electromagnetic potential
$\phi$ is governed by the Schr\"odinger
equation
\begin{equation}\label{eq:schrodinger}
  i\hbar\frac{\partial \psi}{\partial t} + \frac{\hbar}{2 m_e}\nabla^2\psi + e\phi\psi  = 0 ,
\end{equation}
where $\hbar$ is Planck's constant, $m_e$ is the electron mass, and $e$ is the magnitude
of the electron charge. Using the decomposition $\psi = \sqrt{n}\,\exp{iS/\hbar}$ into the amplitude and phase \cite{holland}, (1) can be written as two real equations. Here $n$ is the amplitude and $S$ the phase of the wave function, respectively. Thus, 
Eq.\ (\ref{eq:schrodinger}) becomes
\begin{equation}\label{eq:schrod-cont}
  \frac{\partial n}{\partial t} + \nabla\cdot(n\mathbf{v}) = 0 , 
\end{equation}
and (after taking the derivative of the equation for the phase)
\begin{equation}\label{eq:schrod-mom}
  m_e\left( \frac{\partial}{\partial t} + \mathbf{v}\cdot\nabla\right)\mathbf{v} = e\nabla\phi
    + \frac{\hbar^2}{2m_e}\nabla\left(\frac{\nabla^2\sqrt{n}}{\sqrt{n}}\right) ,
\end{equation}
where the velocity is defined by $\mathbf{v} = \nabla S/m_e$. The last term of Eq.\ (\ref{eq:schrod-mom})
is the gradient of the Bohm--de Broglie potential, and is due to the effect of wave function spreading, giving rise to a dispersive-like term. We also note the
striking resemblance of Eqs.\ (\ref{eq:schrod-cont}) and (\ref{eq:schrod-mom}) to the classical fluid equations.

\subsection{The Pauli description}

Through the Dirac Hamiltonian 
\begin{equation}\label{eq:dirac}
  H = c\mathbf{\alpha}\cdot\left( \mathbf{p} + e\mathbf{A} \right) - e\phi + \beta m_ec^2 ,
\end{equation}
the spin is introduced. Here $\mathbf{\alpha} = (\alpha_1, \alpha_2, \alpha_3)$, $e$ is the magnitude of the electron charge,
$c$ is the speed of light, $\mathbf{A}$ is the vector potential, $\phi$ is the electrostatic potential,
and the relevant matrices are given by
\begin{equation}
  \mathbf{\alpha} = \left( 
    \begin{array}{cc}
      0                 & \bm{\sigma} \\
      \bm{\sigma} & 0
    \end{array}
  \right) , \qquad
  {\beta} = \left( 
    \begin{array}{cc}
      \bm{\mathsf{I}} & 0 \\
      0                     & -\bm{\mathsf{I}}
    \end{array}
  \right) .
\end{equation}
Here $\bm{\mathsf{I}}$ is the unit $2\times 2$ matrix and $\bm{\sigma} = (\sigma_1,\sigma_2, \sigma_3)$, where we have the Pauli spin matrices 
\begin{equation}
\sigma_1 = \left( 
\begin{array}{cc}
0 & 1 \\ 
1 & 0
\end{array}
\right) , \, \sigma_2 = \left( 
\begin{array}{cc}
0 & -i \\ 
i & 0
\end{array}
\right) , \, \text{ and }\, \sigma_3 = \left( 
\begin{array}{cc}
1 & 0 \\ 
0 & -1
\end{array}
\right) .
\label{paulimatrices}
\end{equation}
From the Hamiltonian (\ref{eq:dirac}), the weakly relativistic limit (in $v/c$) gives
\begin{equation}\label{eq:pauli-hamiltonian}
  H = \frac{1}{2m_e}\left( \mathbf{p} + e\mathbf{A} \right)^2 
    + \frac{e\hbar}{2m_e}\mathbf{B}\cdot\mathbf{\sigma} - e\phi .
\end{equation}
We see that the electron's magnetic moment is given by $\mathbf{m} = 
-\mu_B\langle\psi|\mathbf{\sigma}|\psi\rangle/\langle\psi|\psi\rangle$, where 
$\mu_B = e\hbar/2m_e$ is the Bohr magneton, giving a contribution $-\mathbf{B}\cdot\mathbf{m}$
to the energy. The latter shows the paramagnetic property of the electron, where the spin vector is
anti-parallel to the magnetic field in order to 
minimize the energy of the magnetized system.  
Given any operator $F$ and the hamiltonian (\ref{eq:pauli-hamiltonian}), the relation $dF/dt = \partial F/\partial t + (1/i\hbar)[F, H]$,
where $[,]$ is the Poisson bracket, gives the evolution equations for the position and momentum in the Heisenberg picture
\cite{holland}
\begin{equation}
  \frac{d\mathbf{x}}{dt} = \frac{1}{m_e}\left( \mathbf{p} + e\mathbf{A} \right) \equiv \mathbf{v},
\end{equation}
\begin{equation}
  m_e\frac{d\mathbf{v}}{dt} = -e\left( \mathbf{E} + \mathbf{v}\times\mathbf{B} \right) 
    - \frac{2}{\hbar}\mu_B\mathbf{\nabla}(\mathbf{B}\cdot\mathbf{s}) ,
\end{equation}
respectively, while the spin operator satisfies the evolution equation 
\begin{equation}
  \frac{d\mathbf{s}}{dt} = \frac{2}{\hbar}\mu_B \mathbf{B}\times\mathbf{s}, 
\end{equation}
showing the spin precession in an external magnetic field. Here the spin operator is given by $
  \mathbf{s} = ({\hbar}/{2})\bm{\sigma} $.
The above equations thus gives the quantum operator equivalents of the 
equations of motion for a classical particle, including the evolution of the
spin in a magnetic field. 

The non-relativistic evolution of spin-$\tfrac{1}{2}$ particles, as
described by the two-component spinor $\varPsi_{(\alpha)}$, is given by
the Pauli equation (see e.g. \cite{holland}) 
\begin{equation}  \label{eq:pauli}
i\hbar\frac{\partial\psi}{\partial t} + \left[ \frac
{\hbar^2}{2m_e}\left(\nabla + \frac{ie}{\hbar}%
\mathbf{A} \right)^2 - \mu_B\mathbf{B}\cdot\mathbf{\sigma} + e\phi %
\right] \psi = 0 ,
\end{equation}
where $\mathbf{A}$ is the vector potential, 
$\mu_B = e\hbar/2m_e$ is the Bohr magneton,  
and $\mathbf{\sigma} = (\sigma_1, \sigma_2, \sigma_3)$ is the Pauli spin
vector. 

We may, as in the Schr\"odinger case, decompose 
the electron wave function $\psi$ into its amplitude and phase. Thus, we let $\psi = \sqrt{n}\,\exp(iS/\hbar)\varphi$,
where $\varphi$ is a unit two-spinor.
Then we have the conservation equations 
\begin{equation}  \label{eq:pauli-cont}
  \frac{\partial n}{\partial t} + \mathbf{\nabla}\cdot(n\mathbf{v}) = 0
\end{equation}
and 
\begin{eqnarray} 
  m_e\frac{d\mathbf{v}}{d t} = -e (\mathbf{E}
  + \mathbf{v}\times\mathbf{B})  + \frac{\hbar^2}{2m_e}\nabla\left( \frac{\nabla^2\sqrt{n}}{\sqrt{n}} \right)
  - \frac{2\mu_B}{\hbar}({\nabla}\otimes\mathbf{B}%
  )\cdot\mathbf{s}  - \frac{1}{%
  m_en}{\nabla}\cdot\left(n%
  \bm{\mathsf{\Sigma}} \right)
\label{eq:pauli-mom}
\end{eqnarray}
respectively. The spin contribution to Eq.\ (\ref{eq:pauli-mom}) is consistent with the results
of Ref.\ \cite{suttorp}. Here the velocity is defined by 
\begin{equation}
\mathbf{v} = \frac{1}{m_e}\left( {\nabla}S -
i\hbar\varphi^{\dag}\mathbf{\nabla}\varphi \right) + \frac{e\mathbf{A}}{m_ec} ,
\end{equation}
the spin density vector is 
\begin{equation}
\mathbf{s} = \frac{\hbar}{2}\varphi^{\dag}\mathbf{\sigma}%
\varphi ,
\end{equation}
which is normalized according to $
  |\mathbf{s}| = \hbar/2 $,  
and we have defined the
symmetric gradient spin tensor 
\begin{equation}
\bm{\mathsf{\Sigma}} = (\mathbf{\nabla}{s}_a)\otimes(%
\mathbf{\nabla}{s}^a) .
\end{equation}
Moreover, contracting Eq.\ (\ref{eq:pauli}) by $\psi^ {\dag}%
\bm{\sigma}$, we obtain the spin evolution equation 
\begin{eqnarray}  \label{eq:pauli-spin}
  \frac{d\mathbf{s}}{dt} = \left\{ \frac{2\mu_B}{\hbar}\mathbf{B}
    - \frac{1}{m_en}\left[\partial_a(n\partial^a\mathbf{s}) \right]\right\}\times\mathbf{s}.
\end{eqnarray}
We note that the last equation allows for the introduction
of an effective magnetic field $\mathbf{B}_{\rm eff} \equiv 
({2\mu_B}/{\hbar})\mathbf{B} - (m_en)^{-1}\left[\partial_a(n\partial^a\mathbf{s}) \right]$. However, this
will not pursued further here (for a discussion, see Ref.\ \cite{holland}).


\section{Collective plasma dynamics}
As pointed out in the previous section, the route from 
single wavefunction dynamics to collective effects introduces 
a new complexity into the system. At the classical level, the ordinary pressure 
is such an effect. In the quantum case, a similar term, based on the
thermal distribution of spins, will be introduced. 

\subsection{Fluid model}

\subsubsection{Plasmas based on the Schr\"odinger model}
Suppose that we have $N$ electron wavefunctions, and that the total system
wave function can be described by the factorization $\psi(\mathbf{x}_1, \mathbf{x}_2, \ldots \mathbf{x}_N) = \psi_{1}\psi_{2} \ldots \psi_{N}$. For each wave function $\psi_\alpha$, we have a corresponding probability
$\mathcal{P}_\alpha$. From this, we first define $\psi_\alpha = n_\alpha\exp(iS_\alpha/\hbar)$ and follow the steps leading to Eqs. (\ref{eq:schrod-cont}) and (\ref{eq:schrod-mom}). We now have $N$ such equations the wave functions $\{\psi_\alpha\}$. Defining \cite{Manfredi2005}
\begin{equation}\label{eq:meandensity}
  n \equiv \sum_{\alpha = 1}^N \mathcal{P}_\alpha n_\alpha 
\end{equation}
and 
\begin{equation}\label{eq:meanvelocity}
  \mathbf{v} \equiv \langle \mathbf{v}_\alpha \rangle = \sum_{\alpha = 1}^N \frac{\mathcal{P}_\alpha n_\alpha \mathbf{v}_\alpha}{n} ,
\end{equation}
we can define the deviation from the mean flow according to 
\begin{equation}
  \mathbf{w}_\alpha = \mathbf{v}_\alpha - \mathbf{v} .
\end{equation}
Taking the average, as defined by (\ref{eq:meanvelocity}), of Eqs. (\ref{eq:schrod-cont}) and (\ref{eq:schrod-mom}) and using the above variables, we obtain the quantum fluid equation
\begin{equation}\label{eq:cont}
  \frac{\partial n}{\partial t} + \nabla\cdot(n\mathbf{v}) = 0  
\end{equation}
and 
\begin{equation}\label{eq:mom-schrod}
  m_en\left(\frac{\partial}{\partial t} + \mathbf{v}\cdot\nabla\right)\mathbf{v} = en\nabla\phi - \nabla p + \frac{\hbar^2n}{2m_e}\nabla\left\langle \left( \frac{\nabla^2\sqrt{n_\alpha}}{\sqrt{n_\alpha}}\right) \right\rangle ,
\end{equation}
where we have assumed that the average produces an isotropic pressure $p = m_en\langle |\mathbf{w}_\alpha|^2\rangle$
We note that the above equations still contain an explicit sum over the electron wave functions. For typical scale lengths larger than the Fermi wavelength $\lambda_F $, we may approximate the last term by the Bohm--de Broglie potential \cite{Manfredi2005}
\begin{equation}
  \left\langle \frac{\nabla^2\sqrt{n_\alpha}}{\sqrt{n_\alpha}} \right\rangle \approx  
  \frac{\nabla^2\sqrt{n}}{\sqrt{n}} .
\end{equation}
Using a classical or quantum model for the pressure term, we finally have a quantum fluid 
system of equations. For a self-consistent potential $\phi$ we furthermore have
\begin{equation}  
  \nabla^2\phi = \frac{e}{\epsilon_0}(n - n_i) .
\end{equation}

\subsubsection{Spin plasmas}

The collective dynamics of electrons with spin, based on a fluid model, was presented in Ref.\ \cite{marklund-brodin2007}.
For the present discussion, we will follow Refs.\ \cite{marklund-brodin2007,brodin-marklund}. 
Suppose that we have $N$ wave functions for the electrons with  
magnetic moment $ -\mu_B$, and that, as in the case of 
the Schr\"odinger description, the total system
wave function can be described by the factorization $\psi = \psi_{1}%
\psi_{2} \ldots \psi_{N}$. Then the density is defined as in
Eq.\ (\ref{eq:meandensity}) and the average fluid velocity defined by (\ref{eq:meanvelocity}).
However, we now have one further fluid variable, the spin vector, and accordingly we let
$\mathbf{S} = \langle\mathbf{s}_\alpha\rangle $. From this we can define the microscopic
microscopic spin density $\bm{\mathcal{S}}_{\alpha} = \mathbf{s}_{\alpha} 
- \mathbf{S}$, such that $\langle\bm{\mathcal{S}}_{\alpha} \rangle
= 0$.

Taking the ensemble average of Eqs.\ (\ref{eq:pauli-cont}) we obtain the continuity equation
(\ref{eq:cont}), while the ensemble average applied to (\ref{eq:pauli-mom}) yield
\begin{equation}
  m_en\left( \frac{\partial }{\partial t}+\mathbf{v}\cdot {\nabla}\right) %
  \mathbf{v} = -en\left( \mathbf{E}+\mathbf{v}\times \mathbf{B}\right)  
  -\mathbf{\nabla}p 
  + \frac{\hbar^2n}{2m_e}\nabla\left( \frac{\nabla^2\sqrt{n}}{\sqrt{n}}\right) + \mathbf{F}_{\text{spin}}  
  \label{eq:mom-pauli}
\end{equation}
and the average of Eq.\ (\ref{eq:pauli-spin}) gives  
\begin{equation}
  n\left( \frac{\partial }{\partial t}+\mathbf{v}\cdot \mathbf{\nabla}\right) %
  \mathbf{S}= \frac{2\mu_B n}{\hbar }\mathbf{B}\times \mathbf{S}-\mathbf{\nabla}\cdot {%
  \bm{\mathsf{K}}} +\mathbf{\Omega}_{\text{spin}}  
\label{eq:spin}
\end{equation}
respectively. Here the force
density due to the electron spin is   
\begin{eqnarray}
  \mathbf{F}_{\text{spin}} = -\frac{2\mu_B n}{\hbar }(\mathbf{\nabla}\otimes \mathbf{B})\cdot %
  \mathbf{S}-\frac{1}{m_e} {\nabla}%
  \cdot \Big[ n\big(\bm{\mathsf{\Sigma}} + \widetilde{\bm{\mathsf{\Sigma}}}\,\big)\Big]  
-\frac{1}{m_e}\mathbf{\nabla}\cdot \big[n(\mathbf{\nabla}%
  S_{a})\otimes \langle \mathbf{\nabla}\mathcal{S}_{\alpha }^{a}\rangle
  +n\langle \mathbf{\nabla}\mathcal{S}_{\alpha }^a \rangle \otimes (%
  \mathbf{\nabla}S_{a})\big],
\end{eqnarray}
consistent with the results in Ref.\ \cite{suttorp}, 
while the asymmetric thermal-spin coupling is 
\begin{equation}
  \bm{\mathsf{K}} = n\langle \mathbf{w}_{\alpha }\otimes \bm{\mathcal{S}}_{\alpha}\rangle
\end{equation}
and the nonlinear spin fluid correction is 
\begin{eqnarray}
  &&\mathbf{\Omega}_{\text{spin}} = \frac{1}{m_e}\mathbf{S}\times \lbrack \partial
  _{a}(n\partial ^{a}\mathbf{S})]+\frac{1}{m_e}\mathbf{S}\times \lbrack \partial
  _{a}(n\langle \partial ^{a}\bm{\mathcal{S}}_{\alpha }\rangle )]  \notag
\\
&&\quad +\frac{n}{m_e}\left\langle \frac{\bm{\mathcal{S}}_{\alpha}}{%
  n_{\alpha}}\times \left\{ \partial _{a}[n_{\alpha}
  \partial ^{a}(\mathbf{S} + \bm{\mathcal{S}}_{\alpha })]\right\} \right\rangle 
\end{eqnarray}
where $\bm{\mathsf{\Sigma}}=(\mathbf{\nabla}S_{a})\otimes (\mathbf{\nabla}S^{a})$
is the nonlinear spin correction to the classical momentum equation, $%
\widetilde{\bm{\mathsf{\Sigma}}}=\langle (\mathbf{\nabla}\mathcal{S}_{(\alpha
)a})\otimes (\mathbf{\nabla}\mathcal{S}_{(\alpha )}^{a})\rangle $ is a pressure
like spin term (which may be decomposed into trace-free part and trace), and $[(\mathbf{\nabla}%
\otimes \mathbf{B})\cdot \mathbf{S}\,]^{a}=(\partial ^{a}B_{b})S^{b}$. 
Here the indices $a,b,\ldots = 1,2,3$ denotes the Cartesian components 
of the corresponding tensor.
We note that, apart from the additional spin density evolution equation (\ref{eq:spin}), the 
momentum conservation equation (\ref{eq:mom-pauli}) 
is considerably more complicated compared to the Schr\"odinger case represented by (\ref{eq:mom-schrod}).
Moreover, Eqs.\ (\ref{eq:mom-pauli}) and  (\ref{eq:spin}) still contains the explicit sum over the $N$
states, and has to be approximated using insights from quantum kinetic theory or some
effective theory. 

The coupling between the quantum plasma species is mediated by the
electromagnetic field. By definition, we let $\mathbf{H} = \mathbf{B}/\mu_0 - \mathbf{M}$
where $\mathbf{M} = -2n\mu_B\mathbf{S}/\hbar$ is the magnetization due to the spin sources. 
Amp\`ere's law $\mathbf{\nabla}\times\mathbf{H} = \mathbf{j} + \epsilon_0\partial_t\mathbf{E}$ 
takes the form 
\begin{equation}
  \mathbf{\nabla} \times \mathbf{B}=\mu _{0}(\mathbf{j} + \mathbf{\nabla} \times
\mathbf{M}) + \frac{1}{c^2}\frac{\partial\mathbf{E}}{\partial t},  
\label{Eq-ampere}
\end{equation}
where $\mathbf{j}$ is the free current contribution
The system is closed by Faraday's law 
\begin{equation}
  \mathbf{\nabla} \times \mathbf{E}=-\frac{\partial\mathbf{B}}{\partial t} . \label{Eq-Faraday}
\end{equation}

\section{Distribution functions in quantum mechanics}

The Wigner distribution \cite{wigner} function is the quantum version of the classical phase-space distribution function. It was developed during the 20s and 30s following the success of classical non-equilibrium statistical mechanics. However, since in the quantum case the Heisenberg uncertianty principle has to hold, the introduction of a distribution function is not straightforward. There are in fact infinitely many different ways to construct a quantum analogue to the classical distribution function. Except for the Wigner function other popular distribution functions are the Husimi (or Q-) function \cite{Husimi1940} and the Glauber-Sudarshan P-distribution \cite{Glauber1963,Sudarshan1963}. 

There are some properties that are natural to impose on the distribution function: 
\begin{itemize}
	\item Non-negativity, i.e. $f(\mathbf x, \mathbf p) > 0$. 
	\item Correct marginal distributions, i.e. 
	\begin{eqnarray}
		f(\mathbf x) &=& \int d^3 p f(\mathbf x, \mathbf p) \\
		f(\mathbf p) &=& \int d^3 x f(\mathbf x, \mathbf p) , 
	\end{eqnarray}
	should yield, respectively, the probability density to find a particle around the space point $\mathbf x$ or with a momentum around the point $\mathbf p$ in momentum space. 
\end{itemize}
Other properties that can be imposed on the distribution function are that it is real valued, linear in the density matrix and that one can find a complete orthonormal set in the space the distribution function. Since the particle position and momentum cannot be simultaneously known, it is not possible to find a quantum distribution function satisfying all of the conditions above. In order for the description to be complete and equivalent to the density matrix formalism it must be possible to calculate the expectation value of any observable using the distribution function. To do so the operator corresponding to the observable is first mapped to a phase space function $\hat O = O(\hat{\mathbf x}, \hat{\mathbf p}) \rightarrow O(\mathbf x, \mathbf p)$ and the phase space average weighted by the distribution function is them calculated according to 
\begin{equation}
	\langle \hat O \rangle = \int d^3 x d^3 p f(\mathbf x, \mathbf p) O(\mathbf x, \mathbf p) . 
\end{equation}
The mapping from the operator space to phase space depends on which distribution function is used, see \cite{lee} for details. In this proceedings the phase space distribution function we will use is the Wigner distribution function \cite{wigner}. Given the density matrix $\rho$ it is defined by
\begin{equation} \label{wignertrans}
	f_W(\textbf{x}, \textbf{p}) = \frac{1}{(2\pi\hbar)^{3}} \int d^3 y\, 
		e^{ - i\mathbf{p}\cdot\mathbf{y}/\hbar}
		\rho( \mathbf x + \mathbf y/2,  \mathbf x - \mathbf y/2)  .
\end{equation}
It produces the correct marginal distributions, but it may become negative in some regions. The correspondence between operators and phase space functions is in this case is called Weyl correspondence \cite{weyl} and is given by
\begin{equation} \label{weyl}
	\hat O (\hat{\mathbf x}, \hat{\mathbf p}) \leftrightarrow 
	O (\mathbf x, \mathbf p) = \int d^3 y e^{ - i \mathbf p \cdot \mathbf y 
	/\hbar}
	\left< \mathbf x + \mathbf y /2 \right| \hat O \left| \mathbf x - \mathbf y/2 \right> . 
\end{equation}

The discussion so far has been in terms of phase space distribution functions, however, there are similar ways to construct distribution functions for particles with spin \cite{stratonovic, scully83,cohen,chandler}. Here the spin analog to the Q-function will be utilized. It is defined by 
\begin{equation} \label{spintrans}
	 f(\hat{\mathbf s}) = 
	 \mathrm{Tr} \left[   \frac{1}{4\pi} \left( 1 + \hat{ \mathbf{s} }\cdot \bm \sigma \right) \
	 \hat \rho \right] ,
\end{equation}
where $\hat{\mathbf s}$ is a unit vector which will have the same role as the phase space variables $\mathbf x$ and $\mathbf p$, $\bm \sigma$ denotes a vector with the Pauli matrices [Eq.\ (\ref{paulimatrices})] as its components and $\hat \rho$ is the $2 \times 2$ density matrix. Similarly to the phase space case the spin distribution can be used to calculate the expectation value of any operator acting on the spin degree of freedom by integration. 
Since the Pauli matrices satsifies $\sigma_i^2 = I$, for $i = 1,2,3$ where $I$ is the $2 \times 2$ unit vector the only two operators that can act on spin degree of freedom are the unit operator $I$ and the Pauli matrices $\bm \sigma$. The expectation values of these are given by 
\begin{eqnarray}
	\left< I \right> &=& \int d^2 \hat s f(\hat{\mathbf s}) \\ 
	\left< \bm \sigma \right> &=& 3  \int d^2 \hat s \, \mathbf s f(\hat{\mathbf s}) .
\end{eqnarray}
The somewhat peculiar occurrence of a factor 3 can be understood by noting that a pure quantum state is "smeared out" over the whole sphere in contrast to the classical case where a dipole moment may point in a given direction, see \cite{Zamanian-Marklund-Brodin} for details. 

\subsection{Evolution equation} 
The time evolution of the density matrix is given by
\begin{equation}
	i\hbar \frac{\partial \hat \rho}{\partial t}  = \left[ \hat H , \hat \rho \right].  
\end{equation}
For non-relativistic electrons the Hamiltonian is given by Eq.\ (\ref{eq:pauli-hamiltonian}). Taking this as the starting point and using the two transformations Eqs. (\ref{wignertrans}) and (\ref{spintrans}) the evolution equation for the spin distribution function is obtained
\begin{equation} 
\label{eq:full-wigner-equation-2}
\begin{split}
	&
	\frac{\partial f}{\partial t} + \mathbf v \cdot \nabla_{x} f 
	- \left[
		\frac{e}{m_e}(\mathbf{E} + \mathbf{v}\times\mathbf{B})
		+ \frac{\mu_B}{m_e}\nabla_x[(\overrightarrow \nabla_{\hat{ {s} }} 
		+ \hat{\mathbf{s} }) \cdot \mathbf B)
	\right]\cdot\nabla_vf
	- \frac{2 \mu_B}{\hbar} (\hat{ \mathbf{s} }\times \mathbf B)
		\cdot \nabla_{\hat{ {s} }} f
\\ &\quad
	= 
	- \left[
		\frac{e}{m_e}\left(V - \mathbf{v}\cdot\mathbf{A} \right)
		- \frac{\mu_B}{m} \left( \mathbf B \cdot \overrightarrow \nabla_{\hat{ {s} }} 
		+ \hat{ \mathbf{s} } \cdot \mathbf B 
	\right)
	\right]
	\left[
		 \frac{2m_e}{\hbar} \sin\left(\frac{\hbar}{2m}\overleftarrow{\nabla}_x\cdot\overrightarrow{\nabla}_v \right)
		- \overleftarrow{\nabla}_x\cdot\overrightarrow{\nabla}_v
	\right]f
\\ & \qquad
	- \left[ \frac{e}{m} \mathbf A \cdot \overrightarrow \nabla_{x}
		+ \frac{e^2}{m^2}[(\mathbf{A}\cdot\nabla_x)\mathbf{A}]\cdot\nabla_v
		- \frac{2 \mu_B}{\hbar} (\hat{ \mathbf{s} }\times \mathbf B)
		\cdot \overrightarrow \nabla_{\hat{ {s} }} \right]
		\left[
		\cos \left( \frac{\hbar}{2m_e} 
		\overleftarrow \nabla_{x} 
		\cdot \overrightarrow \nabla_{v}\right) - 1
	\right] f .
\end{split}
\end{equation}
where the Coulomb gauge and the variable transformation $m_e \mathbf v = \mathbf p + e \mathbf A(\mathbf x)$ has been used to simplify the equation somewhat. 
In the equation above the function of the operators are defined by their Taylor expansions. 
A more appealing version of the evolution is obtained if the long scale length limit is considered. Neglecting terms of order $\hbar^2$ and higher the evolution equation is written as
\begin{equation}
\begin{split}
	&
	\frac{\partial f}{\partial t} 
	+ \mathbf v \cdot \nabla_{x} f
	-  \left[
	\frac{e}{m_e} \left( \mathbf E + 	\mathbf v \times \mathbf B \right)
	+ \frac{\mu_B}{m_e} \nabla_{x} (\hat{ \mathbf{s} } \cdot \mathbf B) 
	\right] \cdot \nabla_{v} f
	- \frac{2 \mu_B}{\hbar} (\hat{ \mathbf{s} } \times \mathbf B) 
	\cdot \nabla_{\hat{ {s} }} f 
	- \frac{\mu_B}{m_e} 
	[\nabla_{ x} (\mathbf B \cdot \nabla_{\hat{s}})]
	\cdot \nabla_{v} f 	= 0.
\label{eq:semi-classical}
\end{split}
\end{equation}
Note that the last term contains derivatives both with respect to the velocity $\mathbf v$ and the spin $\hat{\mathbf{s}}$. It is interesting to note that we can derive a similar equation on semiclassical grounds. Starting with the distribution function $F(\mathbf x, \mathbf v, \mathbf s)$ and setting the total derivative along particle paths to zero, we obtain
\begin{equation}
	0 = \frac{D F}{Dt} = \frac{\partial F}{\partial t} 
	+ \mathbf v \cdot \nabla_x F 
	+ \frac{\partial \mathbf v}{\partial t} \cdot \nabla_v F 
	+ \frac{\partial \mathbf s}{\partial t} \cdot \nabla_s F   . 
\end{equation}
Now using equations (9) and (10) we obtain 
\begin{equation}
\begin{split}
	&
	\frac{\partial f}{\partial t} 
	+ \mathbf v \cdot \nabla_{x} f
	-  \left[
	\frac{e}{m_e} \left( \mathbf E + \mathbf v \times \mathbf B \right)
	+ \frac{\mu_B}{m_e} \nabla_{x} (\hat{ \mathbf{s} } \cdot \mathbf B) 
	\right] \cdot \nabla_{v} f
	- \frac{2 \mu_B}{\hbar} (\hat{ \mathbf{s} } \times \mathbf B) 
	\cdot \nabla_{\hat{ {s} }} f 	= 0.
\label{eq:semi-classical2}
\end{split}
\end{equation}
This equation has already been studied in \cite{newarticle}. It is there shown to give rise to new oscillation modes due to the anomalous magnetic moment of the electron. A similar equation has also been studied in \cite{cowley} where it is investigated whether spin may be of importance in magnetic confined fusion experiments. The difference between Eq.\ \eqref{eq:semi-classical} and the semi-classical case Eq.\ \eqref{eq:semi-classical2} is due to the fact that the quantum mechanical probability distribution is smeared out compared to the classical distribution function. 

\subsection{Charge and current density} 
To obtain a self-consistent mean-field description of a plasma the evolution equation for the distribution function has to be coupled to Maxwell's equations. This is done by noting that the quantum mechanical operator corresponding to the density is $\hat n (\mathbf x) = \delta(\hat{\mathbf x} - \mathbf x)$ which according to Eq.\ \eqref{weyl} is transformed into the phase space function $n(\mathbf x) = \delta(\mathbf x - \mathbf x')$. Hence the electron charge density is given by
\begin{equation}\label{charge}
	-e n_e (\mathbf x ,t) = -e \int d^3 x' d^3 v d^2 s 
	\delta(\mathbf x - \mathbf x') f(\mathbf x', \mathbf v, \hat{\mathbf s}, t) = -e \int d^3 v 
	f(\mathbf x, \mathbf v , \hat{\mathbf s}, t) . 
\end{equation}
To this one should add the corresponding charge density for the positive species or add a neutralizing homogenous stationary background charge $e n_0$, where $n_0 = \int d^3 x d^3 v d^2 s f$. 
Due to the spin magnetic moment there will be a magnetization current density and the total electron current density is given by \cite{suttorp}
\begin{equation}\label{current}
	\mathbf j_e (\mathbf x, t) = -e \int d^3 v d^2 s 
	\mathbf v f(\mathbf x, \mathbf v, \hat{\mathbf s} ,t) 
	- 3 \mu_B \nabla_x \times \int d^3 v d^2 s \hat{\mathbf s} f(\mathbf x, \mathbf v, 
	\hat{\mathbf s} ,t ) .
\end{equation}
The first term is seen to be the the free current and last term the curl of the magnetization due to the electron spin.

\section{Implications of gauge invariance} 

The definition of the Wigner function \eqref{wignertrans}  is not gauge invariant since it is a function of the gauge dependent canonical momentum rather than the gauge independent kinetic momentum $m_e \mathbf v = \mathbf p + e \mathbf A(\mathbf x)$. The theory above is hence only valid in the Coulomb gauge. It is possible to modify the definition to obtain a gauge independent Wigner function \cite{serimaa86}. 
In principle, there is nothing that prevents us to use a gauge dependent Wigner function as long as care is taken when doing gauge transformations. However, problems may arise when calculating for example the second order moment of the velocity $\left< \hat v_i \hat v_j \right>$. One might then be tempted to calculate
\begin{equation}
	\int d^3 x\, d^3 v\,  v_i v_j f(\mathbf x, \mathbf v, t) 
	= \int d^3 x\, d^3 p\, [p_i + e A_i (\mathbf x) ] 
	[p_j + e A_j (\mathbf x) ] f(\mathbf x, \mathbf p , t) . 
\end{equation}
However, the phase space function which is related to the operator $\hat v_i \hat v_j$ is \textit{not} $[ p_i + e  A_i (\mathbf x) ][p_j + e A_j (\mathbf x) ]$. In order to obtain the right function it is necessary to first put operator $[ \hat p_i + e A_i (\hat{\mathbf x}) ][ \hat p_j + e A_j (\hat{\mathbf x}) ]$ in Weyl-ordering \cite{weyl} and then make the substitution $\hat{\mathbf x} \rightarrow \mathbf x, \hat{\mathbf p} \rightarrow \mathbf p$. This is in general difficult to do since the vector potential is a function of $\mathbf x$. In the current proceedings we are only considering the charge and current densities Eqs.\ \eqref{charge} and \eqref{current} and this problem never arises.

\subsection{A gauge invariant distribution function for the case of spin-extended phase space}

For completeness the fully gauge invariant distribution function is given here. Following Ref.\ \cite{serimaa86} the Wigner matrix is defined by
\begin{equation}
	W^{\rm GI}(\mathbf x, \mathbf v, \alpha, \beta, t) = 
	\frac{1}{(2\pi\hbar)^3} \int d^3 z \exp\left\{-\frac{im_e }{\hbar} \mathbf v \cdot \left[ \mathbf z 
	-e \int_{-1/2}^{1/2} d\tau \mathbf A(\mathbf x+ \tau \mathbf z, t) \right] \right\}
	\rho\left(\mathbf x + \frac{\mathbf z}{2} , \alpha;  
	\mathbf x - \frac{\mathbf z}{2}, \beta \right) . 
\end{equation}
where $\mathbf v$ is the velocity. The explicit dependence of the vector potential in this construction is there to compensate for the phase factor which the wave function acquires under a gauge transformation. Using the spin transformation \eqref{spintrans}, we obtain a fully gauge invariant distribution function. The evolution equation for the gauge invariant distribution function without spin was derived in Ref.\ \cite{serimaa86}. It is straightforward to generalize this equation to include spin with the result
\begin{eqnarray}
	\frac{\partial f^{\rm GI}}{\partial t} + (\mathbf v + \Delta \tilde{\mathbf v} ) \cdot \nabla_x f^{\rm GI} - 
	\frac{e}{m_e} \left[ (\mathbf v + \Delta \tilde{\mathbf v} ) \times \tilde{\mathbf B} + 
	\tilde{\mathbf E} \right] \cdot \nabla_v f^{\rm GI} - 
	\frac{\mu_B}{m_e} \nabla_x [ (\hat{\mathbf s} + \nabla_{\hat s} ) \cdot \tilde{\mathbf B} ] 
	\cdot \nabla_v f^{\rm GI} 
	- \frac{2\mu_B}{\hbar} \left[ \hat{\mathbf s} \times
	\left(  \tilde{\mathbf B} + \Delta \tilde{\mathbf B} \right) \right] \cdot \nabla_{\hat s} f^{\rm GI} = 0, 
	\label{82}
\end{eqnarray}
where we have defined 
\begin{eqnarray}
	&&\!\!\!\!\!\!\!\!\!\!\!\!\!\!\!\!\!\!
	\tilde{\mathbf E} 
	= \int_{-1/2}^{1/2} d\tau \mathbf E \left(\mathbf x + \frac{i\hbar\tau}{m_e} 
	\nabla_v \right) 
	= \mathbf{E}(\mathbf{x}) + \mathbf{e}(\mathbf{x})  
\\ &&\!\!\!\!\!\!\!\!\!\!\!\!\!\!\!\!\!\!
	\tilde{\mathbf B}  
	= \int_{-1/2}^{1/2} d\tau \mathbf B \left(\mathbf x + \frac{i\hbar\tau}{m_e} 
	\nabla_v \right) 
	= \mathbf{B}(\mathbf{x}) + \mathbf{b}(\mathbf{x}) 
\\ &&\!\!\!\!\!\!\!\!\!\!\!\!\!\!\!\!\!\!
	 	\Delta \tilde{\mathbf v} 
		= - \frac{iq\hbar}{m^2} \int_{-1/2}^{1/2} 
	d\tau\, \tau \mathbf B \left( \mathbf x + \frac{i \hbar \tau}{m_e} \nabla_v  \right) \times \nabla_v 
	= 
	- \frac{e\hbar}{m_e ^2}\left[ \mathbf B(\mathbf x) \int_{-1/2}^{1/2} 
	d\tau\, \tau \sin\left( \frac{\tau\hbar}{m_e}\stackrel{\leftarrow}{\nabla}_x\cdot
	 \stackrel{\rightarrow}{\nabla}_v \right)\right] \times \stackrel{\rightarrow}{\nabla}_v 
	 =  - \frac{e\hbar^2}{12 m_e^3}\mathbf{B}(\mathbf{x})\times\stackrel{\rightarrow}{\nabla}_v (\stackrel{\leftarrow}{\nabla}_x\cdot
	 \stackrel{\rightarrow}{\nabla}_v) + \mathcal{O}(\hbar^4)
\\ &&\!\!\!\!\!\!\!\!\!\!\!\!\!\!\!\!\!\!
	\Delta \tilde{\mathbf B} 
	= - \frac{i\hbar}{m_e} \int_{-1/2}^{1/2} d\tau\, \tau 
	\mathbf B \left( \mathbf x + \frac{i\hbar\tau}{m}\nabla_v \right)  
	 \stackrel{\leftarrow}{\nabla}_x\cdot \stackrel{\rightarrow}{\nabla}_v 
	 =
	 \frac{\hbar}{m_e}\mathbf{B}(\mathbf{x})\int_{-1/2}^{1/2}d\tau\, \tau\sin\left( 
	 	\frac{\tau\hbar}{m_e}\stackrel{\leftarrow}{\nabla}_x\cdot 
		\stackrel{\rightarrow}{\nabla}_v
	 \right)\stackrel{\leftarrow}{\nabla}_x\cdot \stackrel{\rightarrow}{\nabla}_v 
	 =  \frac{\hbar^2}{12 m_e^2}\mathbf{B}(\mathbf{x})(\stackrel{\leftarrow}{\nabla}_x\cdot
	 \stackrel{\rightarrow}{\nabla}_v)^2 + \mathcal{O}(\hbar^4) \, , 
\end{eqnarray}	
and 
\begin{eqnarray}
&&
	\tilde{\mathbf E} 
	= \mathbf{E}(\mathbf{x}) \int_{-1/2}^{1/2}d\tau\, \left[\cos\left(  \frac{\tau\hbar}{m}\stackrel{\leftarrow}{\nabla}_x\cdot
	 \stackrel{\rightarrow}{\nabla}_v\right)  -1 \right]
	 = -\frac{\hbar^2}{24m^2}\mathbf{E}(\mathbf{x})(\stackrel{\leftarrow}{\nabla}_x\cdot
	 \stackrel{\rightarrow}{\nabla}_v)^2 + \mathcal{O}(\hbar^4)
\\ &&
	\tilde{\mathbf B}  
	= \mathbf{E}(\mathbf{x}) \int_{-1/2}^{1/2}d\tau\, \left[\cos\left(  \frac{\tau\hbar}{m}\stackrel{\leftarrow}{\nabla}_x\cdot
	 \stackrel{\rightarrow}{\nabla}_v\right)  -1 \right] 
	 =  -\frac{\hbar^2}{24m^2}\mathbf{B}(\mathbf{x})(\stackrel{\leftarrow}{\nabla}_x\cdot
	 \stackrel{\rightarrow}{\nabla}_v)^2 + \mathcal{O}(\hbar^4)
	,
\end{eqnarray}	
so that (cf. (40))
\begin{equation}
\begin{split}
&
	\frac{\partial f^{\rm GI}}{\partial t} + \mathbf v \cdot \nabla_x f^{\rm GI} - 
	\frac{e}{m_e} \left\{ {\mathbf E} + \mathbf v  \times {\mathbf B}  - 
	\frac{\mu_B}{m_e} \nabla_x [ (\hat{\mathbf s} + \nabla_{\hat s} ) \cdot {\mathbf B} ] 
	\right\} \cdot \nabla_v f^{\rm GI} 
	- \frac{2\mu_B}{\hbar} \left( \hat{\mathbf s} \times{\mathbf B} \right) \cdot \nabla_{\hat s} f^{\rm GI} 
\\ & \qquad
	= -  \Delta \tilde{\mathbf v} \cdot \nabla_x f^{\rm GI} 
	+ \frac{e}{m_e} \left\{ {\mathbf e} + (\mathbf v + \Delta \tilde{\mathbf v} ) \times {\mathbf b} 
	+ \Delta \tilde{\mathbf v}  \times {\mathbf B}
	- \frac{\mu_B}{m_e} \nabla_x [ (\hat{\mathbf s} + \nabla_{\hat s} ) \cdot {\mathbf b} ] 
	 \right\} \cdot \nabla_v f^{\rm GI}
	- \frac{2\mu_B}{\hbar} \left[ \hat{\mathbf s} \times
	\left(  {\mathbf b} + \Delta \tilde{\mathbf B} \right) \right] \cdot \nabla_{\hat s} f^{\rm GI} ,
\label{eq:kinetic_split}
\end{split}
\end{equation}
or, to lowest order in $\hbar$,
\begin{equation}
\begin{split}
&
	\frac{\partial f}{\partial t} + \mathbf v \cdot \nabla_x f^{\rm GI} - 
	\frac{e}{m_e} \left\{ {\mathbf E} + \mathbf v  \times {\mathbf B}  - 
	\frac{\mu_B}{m_e} \nabla_x [ (\hat{\mathbf s} + \nabla_{\hat s} ) \cdot {\mathbf B} ] 
	\right\} \cdot \nabla_v f^{\rm GI} 
	- \frac{2\mu_B}{\hbar} \left( \hat{\mathbf s} \times{\mathbf B} \right) \cdot \nabla_{\hat s} f^{\rm GI} 
\\ & \quad
	\approx \frac{\hbar^2}{24m^2}\Bigg[  
		\left(- \frac{e}{m_e} \left\{ 
		\mathbf{E} + \mathbf v\times\mathbf{B}  
	- \frac{\mu_B}{m_e} 
	\nabla_x \left[ (\hat{\mathbf s} + \nabla_{\hat s} ) \cdot \mathbf{B} \right]  
	 \right\} \cdot \stackrel{\rightarrow}{\nabla}_v 
	+ \frac{2\mu_B}{\hbar} \left( \hat{\mathbf s} \times
	 	\mathbf{B}\right) \cdot \nabla_{\hat s} \right) (\stackrel{\leftarrow}{\nabla}_x\cdot
	 \stackrel{\rightarrow}{\nabla}_v)
\\ & \qquad\qquad\qquad
		+ \frac{2e}{m_e}(\mathbf{B}\times\stackrel{\rightarrow}{\nabla}_v) \cdot \stackrel{\rightarrow}{\nabla}_x  
		- \frac{2e^2}{m_e^2}\left[( \mathbf{B}\times\stackrel{\rightarrow}{\nabla}_v )  \times {\mathbf B}\right]\cdot \stackrel{\rightarrow}{\nabla}_v
	 \Bigg] (\stackrel{\leftarrow}{\nabla}_x\cdot
	 \stackrel{\rightarrow}{\nabla}_v) f^{\rm GI} ,
\label{eq:kinetic_split_approx}
\end{split}
\end{equation}
The gauge invariant Wigner function has a modified Weyl correspondence which is well suited for calculating fluid moments. In order to obtain the phase space $O(\mathbf x, \mathbf v)$ function which corresponds to an operator $O(\hat{\mathbf x}, \hat{\mathbf v})$, all products of the operators $\hat{\mathbf x}$ and $\hat{\mathbf v} \equiv [\hat{\mathbf p} + e \mathbf A(\hat{\mathbf x})]/m_e$ are first ordered in a symmetric form using the commutation relation $\hat{\mathbf x}$ and $\hat{\mathbf p}$ and then the substitution $\hat{\mathbf x} \rightarrow \mathbf x$ and $\hat{\mathbf v} \rightarrow \mathbf v$ is taken (details can be found in \cite{serimaa86}). 

The transformation between the gauge-invariant extended Wigner function and the gauge dependent \textit{dito} can be obtained through a kernel function or through an differential operator defined through
\begin{equation}
	f^{\rm GI}(\mathbf{x},\mathbf{v},t) = h(\mathbf{x}, -i(\hbar/m_e)\nabla_v, t)f(\mathbf{x},\mathbf{v},t) ,
\end{equation} 
where 
\begin{equation}
\begin{split}
	h(\mathbf{x}, \mathbf{u},t) 
	& \equiv \exp[i\theta(\mathbf{x},\mathbf{u},t)] \equiv \exp\left[ -\frac{ie}{\hbar}\mathbf{u}\cdot\left( 
		-\mathbf{A}(\mathbf{x},t) + \int_{-1/2}^{1/2}d\tau\,\mathbf{A}(t, \mathbf{x} + \tau\mathbf{u})
	\right) \right]
	= \exp\left[ - \frac{ie}{\hbar}\mathbf{u}\cdot\left( 
		\left.\frac{2\sinh z}{z}\right|_{z = \mathbf{u}\cdot\nabla_x} - 1
	\right)\mathbf{A}(\mathbf{x},t) \right] 
	\\ & = 1 + \left[ \exp(i\theta) - 1 \right] \equiv 1 + \hat{h}(\mathbf{x}, \mathbf{u},t),
\end{split}
\end{equation}
where the part $\hat{h}$ gives the QM correction to between the gauge-invariant and gauge-dependent distribution functions. Thus, expressed through the gauge dependent extended Wigner function, Eq.\ (\ref{eq:kinetic_split}) contains terms proportional to $\hbar^2$ also on the left hand side. Therefore, what is meant with quantum mechanical corrections is dependent on the corresponding choice of Wigner function. We explicitly have
\begin{equation}
	f^{\rm GI}(\mathbf{x},\mathbf{v},t) = \exp\left[ - \frac{e}{m_e} \mathbf{A}(\mathbf{x},t)\left( 
		\frac{2\sin((\hbar/m_e)\stackrel{\leftarrow}{\nabla}_x\cdot\stackrel{\rightarrow}{\nabla}_v)}{(\hbar/m_e)\stackrel{\leftarrow}{\nabla}_x\cdot\stackrel{\rightarrow}{\nabla}_v} - 1
	\right) \cdot\stackrel{\rightarrow}{\nabla}_v\right] f(\mathbf{x},\mathbf{v},t) ,
	= f(\mathbf{x},\mathbf{v},t) -  \frac{e\hbar^2}{24m_e^3}\mathbf{A}(\mathbf{x})\cdot \stackrel{\rightarrow}{\nabla}_v (\stackrel{\leftarrow}{\nabla}_x\cdot\stackrel{\rightarrow}{\nabla}_v)^2 f(\mathbf{x},\mathbf{v},t) + \mathcal{O}(\hbar^4) ,
\end{equation} 
giving us the relation between the gauge independent and gauge dependent distribution functions as an infinite series.

\section{Summary and discussion}\label{sec:sum}

In the present paper we have given an overview of the field of quantum kinetic evolution equations, as given by Eqs. \eqref{eq:full-wigner-equation-2} and (\ref{eq:kinetic_split}), for a
quasi distribution function of electrons, based on a Wigner transformation
of the density matrix, together with a spin operator contracting the $2\times 2$ 
Wigner-matrix to a scalar function  $f(\mathbf{x}, \mathbf{p}, \mathbf{s})$ (or, equivalently, $f(\mathbf{x}, \mathbf{v}, \mathbf{s})$). The free current and the magnetization can be directly computed from the quasi distribution
function, and hence Eq.\ \eqref{eq:full-wigner-equation-2} (or the gauge invariant alternative, Eq.\ \eqref{82}) together with Maxwell's equations with the sources \eqref{charge}, and \eqref{current}, form a closed set. A discussion of the gauge problem was given. 

The gauge independent distribution can be used to derive fluid moments \cite{haas109,haas209}. The resulting equations can be used to investigate for example nonlinear phenomena. However, even in the linear case it is necessary to keep up to the sixth order moment in order to retain the lowest order quantum contribution. Hence, at least for linear analysis the kinetic equation is a better tool.

\section*{Acknowledgments}
MM wishes to thank the participants of the Vlasovia 2009 conference for stimulating discussions. This research is supported by the European Research Council under Constract No.\ 204059-QPQV and  the Swedish Research Council under Contracts No.\ 2007-4422.


\begin{thebibliography}{00}
	\bibitem{wigner} E. Wigner, Phys. Rev. \textbf{40}, 749 (1932).
	\bibitem{weyl} H. Weyl, \textit{Group Theory and Quantum Mechanics} (Dover, New York, 1931).
	\bibitem{groenewold} A. Groenewold, Physica \textbf{12}, 405 (1946).
	\bibitem{moyal} J.E. Moyal, Proc. Camb. Phil. Soc. \textbf{45}, 99 (1949).
	\bibitem{zachos} C. Zachos, D. B. Fairlie, and T. L. Curtright (eds.), \textit{Quantum Mechanics in Phase Space: An Overview with Selected Papers} (World Scientific, 2005). 
	\bibitem{Leonhardt1997}
	U. Leonhardt, \textit{Measuring the Quantum State of Light} 
	(Cambridge University Press, Cambridge, 1997).
	\bibitem{haug-jauho}
	H. Haug and A.-P. Jauho, \textit{Quantum Kinetics in Transport and Optics of Semiconductors} (Springer-Verlag, 2007).
	\bibitem{baym-kadanoff} G. Baym and L. P. Kadanoff, Phys. Rev. \textbf{124}, 287 (1961). 
	\bibitem{kadanoff-baym} L. P. Kadanoff and G. Baym, \textit{Quantum Statistical Mechanics} (Benjamin, New York, 1962). 
	\bibitem{rammer} J. Rammer, \textit{Quantum Field Theory of Non-Equilibrium States} (Cambridge University Press, Cambridge, 2007).
	\bibitem{keldysh1} L. P. Keldysh, Sov. Phys. JETP \textbf{7}, 788 (1958).
	\bibitem{keldysh2} L. P. Keldysh, Sov. Phys. JETP \textbf{20}, 1018 (1965).
	\bibitem{eliezer} S. Eliezer, \textit{Applications of Laser-Plasma Interactions} (Taylor \& Francis, 2009).
	\bibitem{drake} R. P. Drake, \textit{High-Energy-Density Physics} (Springer-Verlag, 2006).
	\bibitem{anderson-etal:2002}
	D. Anderson, B. Hall, M. Lisak, and M. Marklund, Phys. Rev. E \textbf{65}, 046417 (2002).
	\bibitem{marklund:2005}
	M. Marklund, Phys. Plasmas \textbf{12}, 082110 (2005).
	\bibitem{shukla-etal:2006}
	P. K. Shukla, S. Ali, L. Stenflo, and M. Marklund, Phys. Plasmas \textbf{13}, 112111 (2006).
	\bibitem{nshukla:2009}
	N. Shukla, G. Brodin, M. Marklund, P. K. Shukla, and L. Stenflo, Phys. Plasmas \textbf{16}, 072114 (2009).
		\bibitem{shukla-eliasson2009} P.K. Shukla and B. Eliasson, arXiv:0906.4051 (2009).
	\bibitem{glenzer} S. H. Glenzer et al., Phys. Rev. Lett. \textbf{98}, 065002 (2007). 
	\bibitem{kritcher} A. L. Kritcher et al., Science \textbf{322}, 69 (2008). 
	\bibitem{lee-etal} H. J. Lee et al., Phys. Rev. Lett. \textbf{102}, 115001 (2009). 
	\bibitem{alivisatos} A. P. Alivisatos, Science \textbf{271}, 933 (1996).
	\bibitem{haas} F. Haas, Phys. Plasmas \textbf{12}, 062117 (2005),
	\bibitem{manfredi-hervieux} G. Manfredi and P.-A. Hervieux, Appl. Phys. Lett. \textbf{91}, 061108 (2007).
	\bibitem{maier} S. A. Maier, \textit{Plasmonics} (Springer-Verlag, 2007).
	\bibitem{epl} M. Marklund, G. Brodin, L. Stenflo, and C. S. Liu, Europhys. Lett. \textbf{84}, 17006 (2008).
	\bibitem{Brodin-Marklund-Manfredi} G. Brodin, M. Marklund, and G. Manfredi, Phys. Rev. Lett. \textbf{100}, 175001 (2008).
	\bibitem{Zamanian-Marklund-Brodin} J. Zamanian, M. Marklund, and G. Brodin, New J. Phys, in press (2010) (arXiv:0910.5165).
	\bibitem{melrose} D. Melrose, \textit{Quantum Plasmadynamics} (Springer-Verlag, 2008).
	\bibitem{stancil-prabhakar} D.D. Stancil and A. Prabhakar, \textit{Spin Waves} (Springer-Verlag, 2009).
	\bibitem{bertotti-etal} G. Bertotti, I. Mayergoyz, and C. Serpico, \textit{Nonlinear Magnetization Dynamics in Nanosystems} (Elsevier, 2009).
	\bibitem{pethick} 
	C. J. Pethick and and H. Smith, \textit{Bose-Einstein Condensation in Dilute Gases} (Cambridge University Press, 2008).
	\bibitem{marklund-brodin2007} M. Marklund and G. Brodin, Phys. Rev. Lett. \textbf{98}, 025001 (2007).
	\bibitem{brodin-marklund} G. Brodin and M. Marklund, New J. Phys. \textbf{9}, 277 (2007).
	\bibitem{brodin-marklund_pop:2007}
	G. Brodin and M. Marklund, Phys. Plasmas \textbf{14}, 112107 (2007).
	\bibitem{brodin-marklund_pre:2007}
	G. Brodin and M. Marklund, Phys. Rev. E \textbf{76}, 055403 (2007).
	\bibitem{grossman} F. Grossman, \textit{Theoretical Femtosecond Physics} (Springer-Verlag, 2008).
	\bibitem{holland} P. R. Holland, \textit{The Quantum Theory of Motion} (Cambridge University Press, 1995).
	\bibitem{kurtsiefer} C. Kurtsiefer, T. Pfau, and J. Mlynek, \textit{Nature} \textbf{386}, 150-153 (1997). 
	\bibitem{Glauber1963} R. J. Glauber, Phys. Rev. Lett. \textbf{10}, 84 (1963).
	\bibitem{Sudarshan1963}	E. C. G. Sudharshan, Phys. Rev. Lett. \textbf{10}, 277 (1963).
	\bibitem{Husimi1940} K. Husimi, Proc. Phys. -Math. Soc. Japan \textbf{22}, 264 (1940).
	\bibitem{hillery} H. Hillery, R. F. O'Connell, M. O. Scully, and E. P. Wigner, Phys. Rep. \textbf{106}, 121 (1984). 
	\bibitem{lee} H-W. Lee, Phys. Rep. \textbf{259}, 147 (1995).
	\bibitem{styer} D.F. Styer, Am. J. Phys. \textbf{64}, 31 (1996).
	\bibitem{ballentine} L.E. Ballentine, \textit{Quantum Mechanics: A Modern Development} (World Scientific, 1998).
	\bibitem{dauger-etal} D.E. Daugera, V.K. Decyk, and J.M. Dawson, J. Comp. Phys. \textbf{209}, 559 (2005).
	
	\bibitem{ballentine-etal} L. E. Ballentine, Yumin Yang, and J. P. Zibin, Phys. Rev. A \textbf{50}, 2854 (1994).
	\bibitem{Schleich2001} W. P. Schleich, \textit{Quantum Optics in Phase Space} 
	(Wiley, Berlin, 2001).	
	\bibitem{Cohen1986} L.\ Cohen, in \textit{Frontiers of Nonequilibrium Statistical Physics}, 
	eds.\ G.\ T.\ Moore and M.\ O.\ Scully (Plenum, New York, 1986).
	
	\bibitem{harriman} J. E. Harriman, J. Chem. Phys. \textbf{88}, 6399 (1988).
	
	\bibitem{lin-ballentine} W. A. Lin and L. E. Ballentine, Phys. Rev. Lett. \textbf{65}, 2927 (1990).
	
	\bibitem{izrailev} F. M. Izrailev, Phys. Rep. \textbf{196}, 299 (1990).
	
	\bibitem{stockmann} H.-J. St\"ockmann, \textit{Quantum Chaos: An Introduction} (Cambidge University Press, 1999).
		
	\bibitem{stratonovic} R. L. Stratonovic, \textit{Zh. Eksp. Teor. Fiz.} \textbf{58}, 1612 (1970) [\textit{Sov. Phys. -JETP} \textbf{31}, 864 (1970)].
	
	\bibitem{scully83} M. O. Scully, \textit{Phys. Rev. D} \textbf{28}, 2477 (1983).
	\bibitem{cohen} L. Cohen, and M. O. Scully, Found. Phys. \textbf{16}, 295 (1986).
	\bibitem{chandler} C. Chandler, L. Cohen, C. Lee, M. Scully, and K. W\'odkiewicz, \textit{Found. Phys.} \textbf{22}, 867 (1992). 
		\bibitem{carruthers} P. Carruthers and F. Zachariasen, Rev. Mod. Phys. \textbf{55}, 1 (1983).
	\bibitem{scully} M. O. Scully and K. W\'odkiewicz, Found. Phys. \textbf{24}, 85 (1994). 
	\bibitem{radcliffe} J. M. Radcliffe, J. Phys. A \textbf{4}, 313 (1971). 
	\bibitem{arecchi} F. T. Areccchi, E. Courtens, R. Gilmore, and H. Thomas, Phys. Rev. A \textbf{6}, 2211 (1971). 
	\bibitem{newarticle} G. Brodin, M. Marklund, J. Zamanian, {\AA}. Ericsson, and P. L. Mana, Phys. Rev. Lett. \textbf{101}, 245002 (2008).
	\bibitem{arnold} A. Arnold and H. Steinrück, ZAMP \textbf{40}, 6 (1989).
	\bibitem{oconnell} R. F. O'Connell and E. P. Wigner, Phys. Rev. A \textbf{30}, 5 (1984).
	\bibitem{manfredi2008} G. Manfredi, P. A. Hervieux, Y. Yin, N. Crouseilles, \textit{Advances in the atomic-scale modeling of nanosystems and nanostructured materials, Lecture Notes In Phyiscs}, eds.\ C. Massobrio, H. Bulou and C. Goyenex (Springer, Heidelberg, 2009).
	
	 \bibitem{serimaa86} O. T. Serimaa, J. Javanainen, and S. Varr\'o, Phys. Rev. A \textbf{33}, 2913 (1986).  
	 \bibitem{bonitz} M. Bonitz \textit{Quantum Kinetic Theory} (B. G. Tubner Stuttgart - Leipzig, 1998). 
	 \bibitem{suttorp} S. R. De Groot, and L. G. Suttorp, \textit{Foundations Of Electrodynamics}, (North-Holand Publishing Company - Amsterdam, 1972). 
	 \bibitem{handbook}
	 M. Abramowitz and I. A. Stegun (eds.), \textit{Handbook of Mathematical Functions with Formulas, Graphs, and Mathematical Tables} (Dover Publications, 1972).
	 \bibitem{bogolyubov} N. N. Bogolyubov \textit{Zh. Eksp. Teor. Fiz.} \textbf{16}, 691 (1946); 
	[\textit{JETP} \textbf{10}, 265 (1946)]. 
	\bibitem{swanson} D. G. Swanson, \textit{Plasma Waves} (Taylor \& Francis, 2003). 
	\bibitem{fortov} V. Fortov, I. Iakubov, and A. Khrapak, \textit{Strongly Coupled Plasma} (Oxford University Press, New York, 2006).
	\bibitem{cowley} S. C. Cowley, R. M. Kulsrud and E. Valeo, Phys. Fluids. \textbf{29}, 430 (1986).
	\bibitem{demircioglu} B. Demircioglu, and A. Vercin, Ann. Phys. \textbf{305}, 1 (2003).
	\bibitem{landau} L. D. Landau, E. M. Lifshitz, \textit{Quantum Mechanics: Non-relativistic Theory} (Butterworth-Heinemann, 1981).   
	\bibitem{wigner71} E. Wigner, \textit{Perspectives in Quantum Theory}, eds.\ W. Yourgrau and, A. van der Merwe (Dover, New York, 1971).  
	\bibitem{mizrahi} S. S. Mizrahi, Physica A \textbf{150}, 541 (1988). 
	\bibitem{Manfredi2005} G. Manfredi, Fields Institute Communications Series. \textbf{46}s, 263-287 (2005) (quant-ph/0505004).
	\bibitem{haas109} F. Haas, M. Marklund, G. Brodin, and J. Zamanian, \textit{Phys. Lett. A} \textbf{374}, 481 (2009). 
	\bibitem{haas209} F. Haas, J. Zamanian, M. Marklund, G. Brodin, arxiv:0912.4718v1 (2009). 
\end{thebibliography}
\end{document}